\documentstyle[twoside,fleqn,espcrc2,epsf]{article}

\def\la{\raise.16ex\hbox{$\langle$} } 
\def\ra{\raise.16ex\hbox{$\rangle$} } 
\def\psibar{ \psi \kern-.65em\raise.6em\hbox{$-$} }
\def\plaq{\Box}
\newcommand{\be}{\begin{equation}}
\newcommand{\ee}{\end{equation}}
\newcommand{\bea}{\begin{eqnarray}}
\newcommand{\eea}{\end{eqnarray}}

\newcommand{\beq}{\begin{equation}}
\newcommand{\eeq}{\end{equation}}

\newcommand{\AmS}{{\protect\the\textfont2
  A\kern-.1667em\lower.5ex\hbox{M}\kern-.125emS}}

\title{Smeared Gauge Fixing}

\author{J.E. Hetrick\address{Physics Department, 
University of the Pacific, Stockton, CA 95211}%
        and 
        Ph. de Forcrand\address{Swiss Center for Scientific Computing,
ETH-Z\"urich, CH-8092 Z\"urich, Switzerland}}
       
\begin{document}

\begin{abstract}
We present a new method of gauge fixing to standard lattice Landau
gauge, Max Re Tr $\sum_{\mu,x}U_{\mu,x}$, in which the
link configuration is recursively smeared; these smeared links are
then gauge fixed by standard extremization. The resulting gauge
transformation is simultaneously applied to the original links.
Following this preconditioning, the links are gauge fixed again as
usual. This method is free of Gribov copies, and we find that for physical
parameters ($\beta \geq 2$ in SU(2)), it generally results in the gauge
fixed configuration with the globally maximal trace.
This method is a general technique for
finding a unique minimum to global optimization problems.
\end{abstract}

\maketitle


There are two outstanding difficulties with gauge fixing on the lattice:
\begin{enumerate}
\item {\it The Gribov problem}: covariant gauge fixing inevitably 
leads to multiple solutions of the gauge fixing condition.
\item {\it The Smoothness condition}: if one alters the gauge
condition so that it is free of Gribov copies (like axial gauge) it is
generally not smooth, and difficult to compare to perturbation theory.
\end{enumerate}

Below we present a method addressing these two issues which is
both simple and fast. It was inspired by the use of improved operators
and the smoothing properties of smearing links.  It is
free of Gribov copies, and usually produces the smoothest (Max
$I(U_\mu,G) \equiv {\rm Re~Tr}~\sum_{x,\mu} U^G_\mu$) configuration.

It should be contrasted to other covariant solutions to the gauge
fixing problem \cite{LAP}, which are also Gribov copy free, but
require non-trivial computational resources and do not rotate fields
to Landau gauge.

\section{Method}

We begin by iteratively smearing a copy of the links, ie. recursively
smearing again the smeared links, with an APE type smearing process.
In this study we use the method of \cite{MILCSMEAR} in which links are
averaged with connecting staples; staples have weight $w$, and the
resulting link is reunitarized into SU(N).

We have found numerically that as we iterate this process there is a
critical weight $w_c$ below which the configuration is cooled, ie. the
average plaquette, moves closer and closer to the identity; with a
weight above $w_c$ smearing moves the links to a ``rough''
configuration with Tr$\la\plaq\ra \sim 0$.  The value of this critical
weight seems to depend weakly on $\beta$ and volume: $w_c \sim
0.5-0.6$.

This method of cooling however is {\it equivariant}: 
{If starting configurations $U^A_\mu$ and $U^B_\mu$ are related by
a gauge transformation $G^{AB}$, then the corresponding smeared
configurations $V^A_\mu$ and $V^B_\mu$ are related by
the same gauge transformation 
$G^{AB}$}. This is useful
since the smeared configuration is approaching the trivial orbit,
where we know the unique Landau gauge configuration. It is the
configuration in which each link is constant and diagonal: $U_\mu(x) =
(D_\mu)^{1/n_\mu}$ such that ${\rm Tr} D_\mu = {\rm Tr} P_\mu$ where
$P_\mu$ is the smeared Polyakov loop in the $\mu$-direction (made of
$n_\mu$ links). We neglect here issues of degeneracy of eigenvalues,
etc.

Thus if we smear extremely close to the trivial orbit, we can find a
$G_{VD}$ which rotates the smeared links $V_\mu$ to Landau gauge
uniquely, then apply this $G_{VD}$ to the original links $U_\mu$,
producing a unique configuration $D^\prime_\mu$; due to the
nonlinearity of non-abelian gauge transformations, it is neither
diagonal nor satisfies the Landau gauge condition in SU(N)
theory. Still it represents a unique starting point on the physical
orbit, from which we can gauge fix to Landau gauge and be assured of a
unique solution. This is displayed in figure 1.
\begin{figure}[htb]
\vspace{-1cm}
\centerline{
\epsfxsize=9cm 
\epsffile[25 170 500 525]{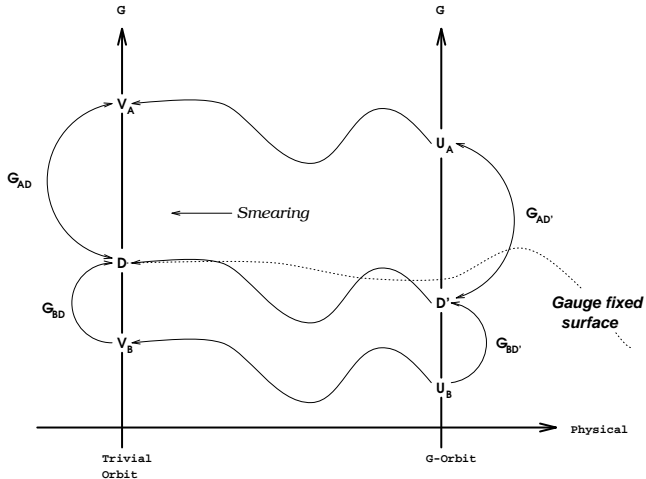}
 }
\vspace{-1cm}
\caption{Gauge fix $V_\mu$ to $D_\mu$ and use the same $G$ to fix
$U_\mu$ to $D^\prime_\mu$.}
\vspace{-1cm}
\label{FIG1}
\end{figure}

\section{A simpler way}

In our experiments we have found that it is not necessary to smear the
links all the way to the trivial orbit; after only a moderate amount
of smearing, gauge fixing of the smeared links is unique. Our view of
this is captured below in an {\em artist's rendering} of the change in
$I(V,G)$ as we smear.
\begin{figure}[htb]
\centerline{
\epsfxsize=8.4cm 
\epsffile[130 295 500 535]{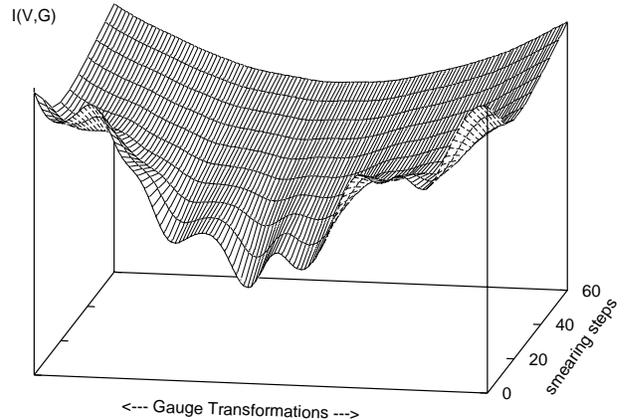}
 }
\vspace{-0.8cm}
\caption{The gauge fixing functional $-I(V,G)$ changes as we smear 
$V_\mu$. Initially the functional has many relative minima leading to
Gribov copies. As the links are recursively smeared $-I(V,G)$ becomes
smoother, developing a unique minimum.}
\vspace{-0.2cm}
\label{FIG2}
\end{figure}

If the original configuration is smooth enough, the global
maximum on the smeared orbit may lie in the basin of attraction of the
(gauge transformed) global maximum on the physical orbit. When this
happens, the preconditioning step of gauge fixing the smeared links to
Landau gauge leads to the final copy being the one of maximal
$I(U_\mu,G)$, ie. the smoothest copy. If the original configuration is
quite rough (unphysically so), there often is a mismatch between the global
maxima of the smeared and original $I$ functions. The resulting
configuration then does not have the globally maximal trace,
although the trace is usually still rather high among the copies. In either
event though, {\bf gauge fixing (to Landau gauge) is unique}.

\section{Parameters}

In the original form of this method, we should smear until we are
within machine zero of the trivial orbit ($F_{\mu\nu} = 0$), however
we have found that only partial smearing is necessary. The optimal
parameters specifying this have emerged rather heuristically, and we
will simply summarize them here and refer the reader to our more
complete paper on this method of gauge fixing \cite{HdeF}. Also we
have studied small lattices so far, up to $12^4$, and we gauge
fix to variations smaller that $10^{-13}$ in $I$.

The smearing coefficient $w$, which is the weight given to the staples
should be as large as possible, but below the critical value so that
the smeared configuration is smoothed. We have found that $w \sim 0.3$
works best, looking at how topological objects (which slow the
cooling) are annihilated.

The most important parameter is the number of smearing sweeps done. We
have found that this depends on the number of topological objects in
the configuration. Thus smearing is pursued until the action falls
below the one-instanton bound. On our $8^4$ lattices, we smeared until
the average plaquette was within $10^{-5}$ of {\bf 1}, which then leads
to unique gauge fixing. In figure 3 we show the cooling history of 6
lattices at a range of $\beta$ values. The plateaux show the stability
of instanton pairs under smearing (like under cooling).
\begin{figure}[htb]
\epsfxsize=7cm 
\epsfysize=6cm 
\epsffile[100 254 541 597]{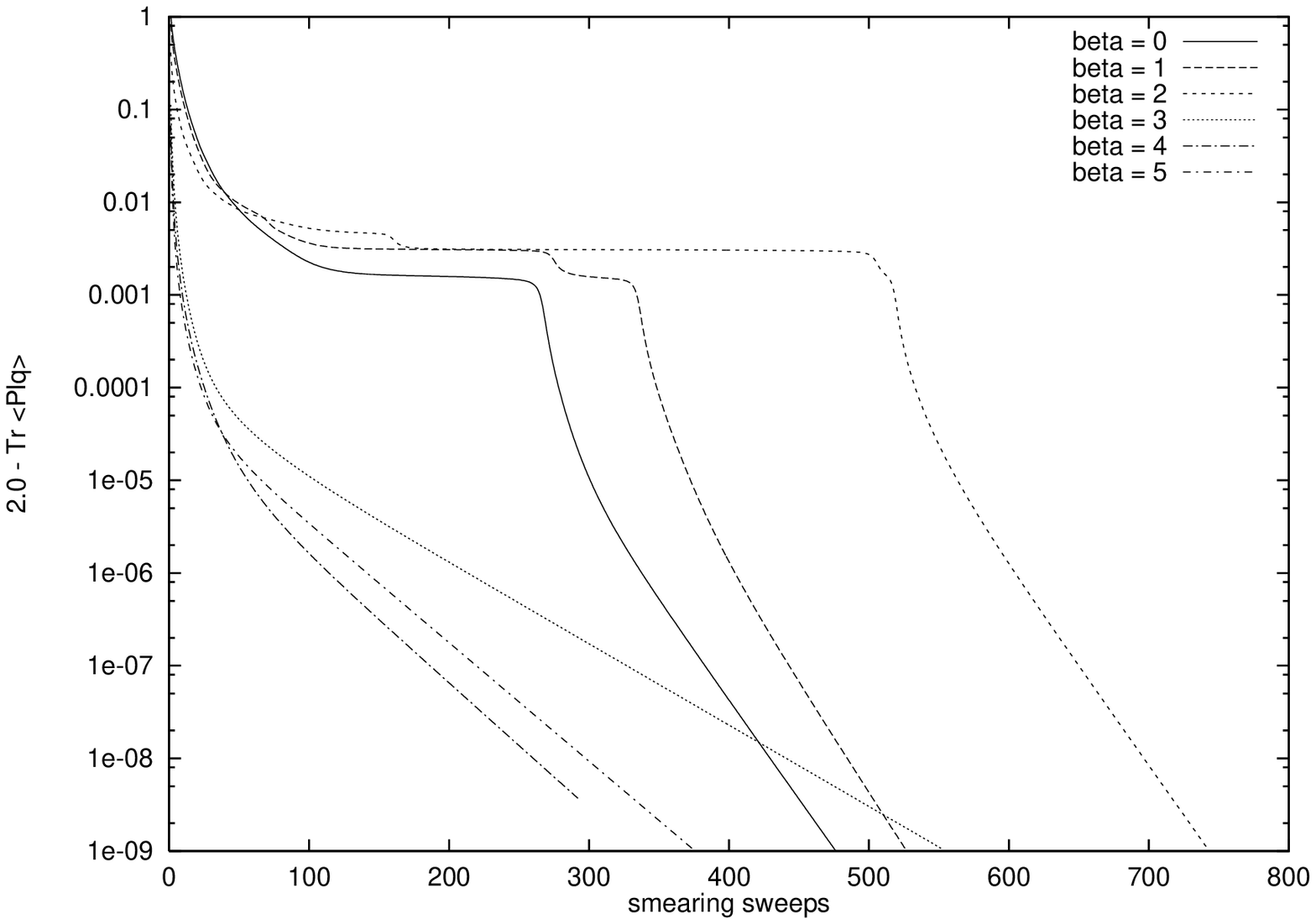}
\vspace{-0.7cm}
\caption{$2.0 - {\rm Tr}\la\plaq\ra$ versus smearing 
($w = 0.3$) on lattices of size $8^4$.}
\label{FIG6}
\end{figure}

\section{Summary}

In figure 4 we display typical results on 100 random gauge
transformations at 3 $\beta$ values: 2.0 (top), 1.75 (middle), and 1.5
(bottom). The graphs contain histograms of the final value of $I$ on
the gauge fixed copies obtained using local extremization, and the arrow
shows the unique copy obtained with our method.
\begin{figure}[ht]
\epsfxsize=8cm 
\epsffile[75 53 542 724]{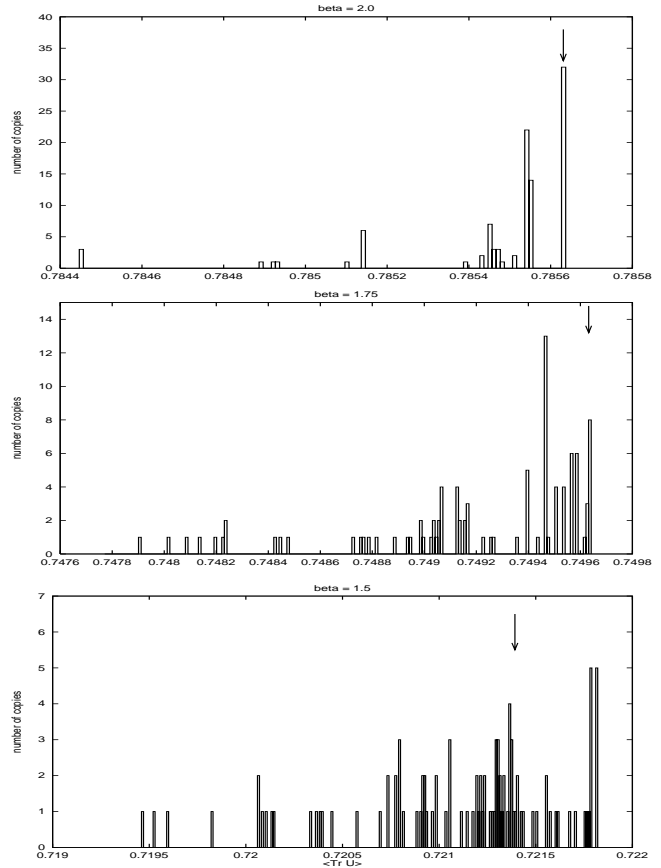}
\vspace{-0.5cm}
\caption{Gribov copies obtained by standard extremization
and the {\em unique copy} $\downarrow$ obtained from smearing.}
\label{FIG4}
\end{figure}
To summarize our method:
\begin{itemize}
\item Recursively smear a configuration $U_\mu$ into
$V_\mu$. 
\item Gauge fix $V_\mu$ by extremizing, 
$I(V,G)\equiv \sum_{x,\mu} {\rm Re~Tr}~ G^\dagger(x) V_\mu(x) G(x+\mu)$,
and apply the same gauge transformation to $U_\mu$.
\item Finally, gauge fix $U_\mu$ as usual, by extremizing $I(U,G)$.
\end{itemize}

This construction trivially generalizes to Coulomb or maximal Abelian gauges,
and can be more generally applied to some optimization problems.

\end{document}